\documentclass{article}
\usepackage{graphicx} 
\usepackage{xcolor}
\usepackage{subcaption}

\usepackage{graphicx} 
\usepackage{amsmath}
\usepackage{float}   
\usepackage{xcolor}
\usepackage{subcaption}
\usepackage{mathtools}
\usepackage[utf8]{inputenc}
\usepackage[backend=biber,style=numeric,sorting=none]{biblatex}
\usepackage[colorlinks=true,citecolor=blue]{hyperref}

\usepackage{amsthm}
\usepackage{amssymb}
\usepackage{authblk}

\title{Nonlinear oscillations of the amplitude of energetic-particle induced geodesic acoustic modes}
\author{E. Sida$^1$, A. Biancalani$^{1,*}$, A. Bottino$^2$, F. Salvarani$^{1,3}$, and R. Wu$^1$\\
{\small
$^1$ De Vinci Higher Education, De Vinci Research Center, 92400 Paris, France\\
$^2$ Max Planck Institute for Plasma Physics, 85748 Garching, Germany\\
$^3$ Department of Mathematics ``F. Casorati'', University of Pavia, 27100 Pavia, Italy\\
$^*$ Contact of corresponding author: alessandro.biancalani@devinci.fr
}}

\begin{document}

\maketitle

\begin{abstract}
Energetic particle induced geodesic acoustic modes (EGAMs) are axisymmetric perturbations of the radial electric field in tokamak plasmas. They are driven unstable by the phase space nonuniformity of a population of energetic particles (EP). In this paper, the nonlinear oscillation in the amplitude of the energetic-particle induced geodesic acoustic modes is studied by means of the gyrokinetic particle-in-cell code ORB5. Similarities are discussed with the beam-plasma instability (BPI), where a Langmuir wave is driven unstable by phase space nonuniformity of a population of energetic electrons. A similar scaling of the nonlinear oscillation frequency as a function of the mode amplitude is found for the EGAMs and for the BPI, confirming that their nonlinear dynamics is strongly determined by the same physical mechanisms. As a product of this study, a novel diagnostics is proposed for the evaluation of the EGAM intensity in tokamak plasmas.
\end{abstract}

\section{Introduction}
Plasmas in magnetic confinement fusion devices are expected to have a large population of energetic (i.e. suprathermal) particles (EP) due to the nuclear fusion reactions, and to the external heating mechanisms. A large variety of instabilities can take energy out of the EP population, via inverse Landau damping~\cite{Landau46}, for example Alfv\'en waves~\cite{Chen16}.  We are interested here in the formation of unstable branches of the geodesic acoustic mode (GAM)~\cite{Winsor68}, due to inverse Landau damping with EPs, giving rise to EP induced GAMs (EGAMs)~\cite{Fu08}. GAMs are a type of zonal (i.e. axisymmetric) flows (ZF), associated to a perturbation of the radial electric field. GAMs are often observed in the presence of turbulence in tokamaks. EGAMs can redistribute the EP population in phase space, thus modifying their distribution function and consequently affecting the heating mechanism.

EGAMs have been studied theoretically~\cite{Fu08,Qiu10,Qiu11,WangPRL13,Zarzoso13,Miki15,Zarzoso17,Sasaki17scirep,DiSiena18,Qiu18,Novikau20,Vannini22,Rettino23} and experimentally (see for example Ref.~\cite{Horvath16}). A role as possible mediators of EPs and turbulence has also been proposed~\cite{Zarzoso13,Sasaki17scirep}. Similarities of the physics of EGAMs with the Beam Plasma Instability (BPI) have been proposed for example in Ref.~\cite{Qiu2011}. The BPI comes from the interaction of a beam of energetic electrons with a Langmuir wave~\cite{Levin72} (see also Ref.~\cite{Carlevaro15} for a recent study). Following the analytical study done in Ref.~\cite{Qiu2011}, numerical simulations have also been performed, and their nonlinear saturation levels due to wave-particle interaction, have been discussed and compared with the BPI~\cite{Biancalani2017,Biancalani2018}.

In this paper, we focus on the EGAM behavior just after the nonlinear saturation. Our goal is to give insights on the amplitude evolution in time, and show how the physics of the BPI can help understanding it. Therefore, we first study the BPI following a model presented by Levin in 1972, see Ref.~\cite{Levin72}. Here the equations of motion are coupled with an energy-balance low for the amplitude of the wave. This model is shown in Sec.~\ref{sec:BPI-model}. In Sec.~\ref{sec:BPI-results} we study the model numerically, implementing a particle-in-cell code. This code is used to investigate the evolution of the amplitude of the BPI. We recover the results of Levin, and in particular we investigate in detail the oscillations during the nonlinear phase.



After having discussed the nonlinear oscillations of the BPI, we perform simulations of EGAMs considering the case of Ref.~\cite{Biancalani2017,Biancalani2018}, where the EGAM had already been studied in detail for its linear phase and nonlinear saturation. Our numerical simulations are performed with the ORB5 particle-in-cell code. ORB5 is a multispecies global gyrokinetic code~\cite{ORB5}. This code is described in Sec.~\ref{sec:model}, where the tokamak case is also shown, and the linear dynamics is given.

A scan with the EP concentration is performed, allowing us to investigate the dynamics of the EGAMs for different drive intensities.
In Sec.~\ref{sec:nonlinear} we discuss the behavior of the maximum of the radial electric field, and we show that during the nonlinear phase, some oscillations of the EGAM radial electric field are present. The scaling of this nonlinear frequency with the mode amplitude is studied. 

Sec.~\ref{sec:comparison}, is devoted to a comparison with BPI and the EGAM. To this aim,  we use the theoretical result of Ref.~\cite{Qiu2011}, that links the bounce frequency of EP trapped in the wave, with the saturation level. Finally, a discussion and outlook is given in Sec.~\ref{sec:conclusion}, where a novel diagnostics is also proposed for the measurement of the EGAM amplitude in tokamak plasmas.

\section{The Beam Plasma Instability model}
\label{sec:BPI-model}

We adopt the model for the dynamics of the beam-plasma instability (BPI) provided by Levin in Ref.~\cite{Levin72}. The BPI comes from the interaction of a beam of electrons and a Langmuir wave. A homogeneous plasma is considered, and the dynamics is studied in the electrostatic limit. Thus, the problem reduces to a 1 dimension in real space (with coordinate $x$), and 1 dimension in velocity space (with coordinate $v$). Consistently with Ref.~\cite{Levin72}, we want to study the amplitude of the Langmuir wave, but we are not interested in the evolution of its frequency and radial structure. Therefore in this model the wave has constant frequency and radial structure (a sinusoidal shape with wavelength $\lambda$). On the other hand, the evolution of the wave amplitude is studied self-consistently with the electron beam nonlinear dynamics. The dynamics of these energetic particles is studied in a self-consistent way, by following their trajectories in the wave field.

The model equations that we use are those of Levin~\cite{Levin72}. The electric field has the form: $E(x,t) = E(t)\sin(kx-\omega t)$, where $k$ and $\omega$ are supposed to be known. The system of equations that rules the dynamics of particles and the evolution in time of the amplitude of the wave is the following:
\begin{equation}
\left\{
\begin{aligned}
\frac{d\xi}{dt} &= v' \\
\frac{dv'}{dt} &= - \frac{e}{m}E(t)\sin(k\xi) \\
\frac{1}{4\pi}E(t)\frac{dE}{dt}
&= -eE(t)\frac{1}{\lambda}
\int_{-\frac{\lambda}{2}}^{\frac{\lambda}{2}}
\int_{-v_0^m}^{v_0^m}
\sin(k\xi)\left(v'+\frac{\omega}{k}\right)
f\left(v'+\frac{\omega}{k}\right)
\, d\xi \, dv
\end{aligned}
\right.
\label{eq:first}
\end{equation}

Here $\xi$ and $v'$ are, respectively, the position and velocity of particles in the wave frame: $\xi = x-\frac{\omega}{k}t$ and $v' = v-\frac{\omega}{k}$.\\ 
To derive the equation for the variation of the power of the field over time the author used a result of O'Niel, (on Ref. \cite{ONeil1965}) stating that the only particles exchanging energy with the wave were the ones whose velocity is close to the resonant velocity of the wave. Therefore, we don't need to integrate in all the real space, but it is sufficient to do the integration in a small interval around the resonance, with boundaries $-v_0^m$ and $v_0^m$. Note that these boundaries are free to choose: the only condition is to take the interval sufficiently large, so that the main physics is all inside, and the result does not depend anymore on their position. For more detail of this model, the reader is referred to Ref.~\cite{Levin72}.

\section{Numerical Simulations of the BPI system}
\label{sec:BPI-results}
As a first step we recovered Levin's numerical results by solving the dimensionless system~\cite{Levin72}:

\begin{equation}
 \left\{
\begin{aligned}
\frac{d\nu}{d\tau} &= -\frac{\mathcal{E}}{2\pi}sin(2\pi\zeta) \\
\frac{d\zeta}{d\tau} &= \nu \\
\frac{d\mathcal{E}}{d\tau} &= 
  16\pi\int_0^{\frac{1}{2}}\int_{-v_0^m}^{v_0^m}\nu_0sin(2\pi\zeta)d\nu_0d\zeta_0
\end{aligned}
\right.   
\end{equation}
This was obtained from system \eqref{eq:first} by defining the dimensionless variables:
\[
\zeta = \frac{1}{2\pi}k\xi \quad \nu = \frac{1}{2\pi}\frac{kv'}{\gamma_L} \quad \mathcal{E} = \frac{eEk}{m\gamma_L^2} \quad \tau = \gamma_L t
\]

After the change of variables, we used Liouville’s theorem that states that the distribution function f remains constant following the particles trajectories. The evolution in time of the dimensionless amplitude is obtained by linearizing the system and taking a Taylor expansion of the first order around $v_{\phi} = \frac{\omega}{k}$ . Then, assuming that $\nu$ and $\xi$ are odd functions, we get:

\[
\frac{d\mathcal{E}}{d\tau} = (\frac{8 \pi^2e^2}{m\gamma_L^2})(\frac{2\pi\gamma_L}{k})(2v_\phi\frac{\partial f_0}{\partial v_\phi})\int_{0}^{\frac{1}{2}}\int_{-\nu_0^m}^{\nu_0^m} sin(2\pi\zeta)\nu_0 = \]
\[ (\frac{16\pi}{\gamma_L})\underbrace{(\frac{2\pi^2e^2}{mk}v_\phi\frac{\partial f_0}{\partial v_\phi})}_{\gamma_L} \int_{-\nu_0^m}^{\nu_0^m} sin(2\pi\zeta)\nu_0
\]
\[
\Rightarrow \frac{d\mathcal{E}}{d\tau} = 16 \pi \int_{-\nu_0^m}^{\nu_0^m} sin(2\pi\zeta)\nu_0
\]
with $f_0$ being the equilibrium distribution function. Here, we have used the expression for the linear growth rate for the inverse Landau Damping: $\gamma_L = \frac{2\pi^2e^2}{mk}v_\phi\frac{\partial f_0}{\partial v_\phi}$. 
 The result is a ODE system that can be solved numerically, with a Runge-Kutta method.
 
\begin{figure}[H]
    \centering
    \includegraphics[width=0.6\textwidth]{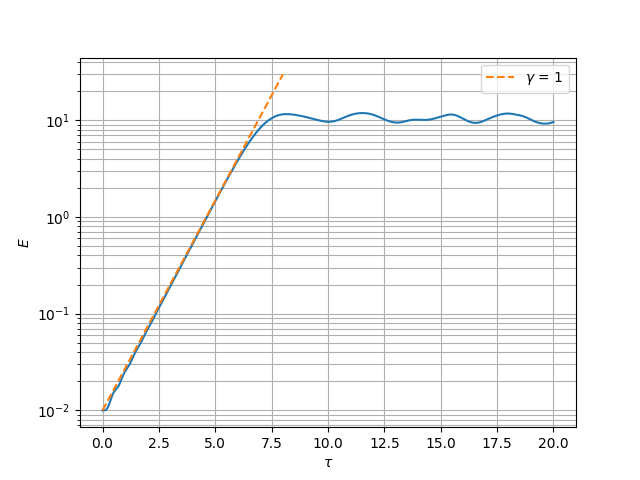}
    \caption{Dimensionless amplitude of the BPI electric field, as a function of $\tau$. After an initial linear phase, the wave saturates. The oscillations that we see during the nonlinear phase have the same frequency of the bounce of the trapped particles, as predicted by Levin. The orange line is obtained with the predicted dimensionless growth rate.}
    \label{fig:el_field}
\end{figure}

 Following Levin's steps we performed a simulation with 4000 particles with $\nu_0 \in (-2,2), \quad E_0 = 0.01$. We remark that the system does not depend on the initial distribution anymore, since the dependence on the initial distribution function is contained in the linear growth rate,  so we can initialize the particles with a uniform distribution. A linear growth of this instability can be seen in the initial phase of the simulation (see Fig.~\ref{fig:el_field}). The growth rate measured matches well the theoretical growth rate: in this case, $\gamma=1$, as the system is normalized to the linear growth rate. A nonlinear saturation of the electric field amplitude at $E\simeq 10$ is observed around $\tau=7.5$. Then, the field amplitude oscillates around this value.

\begin{figure}[h]
    \centering
    \vspace{0.2 cm}
     \includegraphics[width=0.8\textwidth]{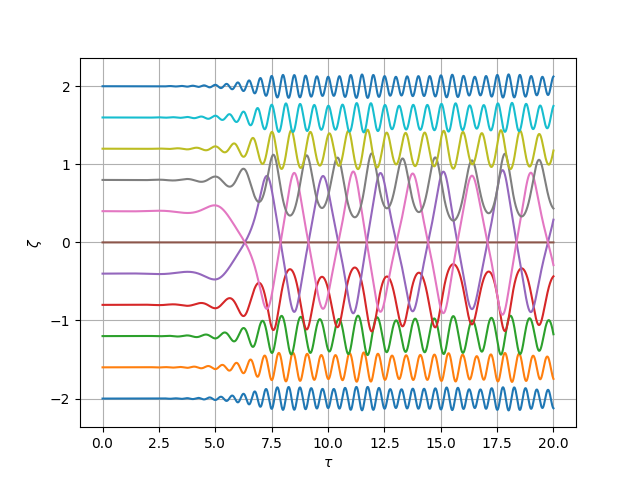}
        \caption{Velocity of particles as a function of $\tau$. The linear and nonlinear phase are clearly visible, and the effect of the field on both passing and trapped particles can be observed in their velocity $\zeta$.}
        \label{fig:velocity}
\end{figure}

It is interesting to study the dynamics of the electron beam during the linear and nonlinear phase. This is shown in Fig.~\ref{fig:velocity}. All particles follow a straight line in the very beginning of the simulation, when the BPI field is negligible. After a while, some particles remain passing, and some become trapped. It is clear the distinction between trapped particles and passing particles: the particles near the resonance velocity are trapped, and in fact their velocity oscillates around zero in $\zeta$. We can also see as the particles in green after an initial phase where they are passing, as the amplitude grows become trapped. On the contrary in violet we see particles that during the saturation phase become passing particle. Then for those whose initial velocity is greater than 1 in module we see that for the entire simulation they aren't trapped.
  
\begin{figure}[H]
    \centering
         \includegraphics[width=0.8\textwidth]{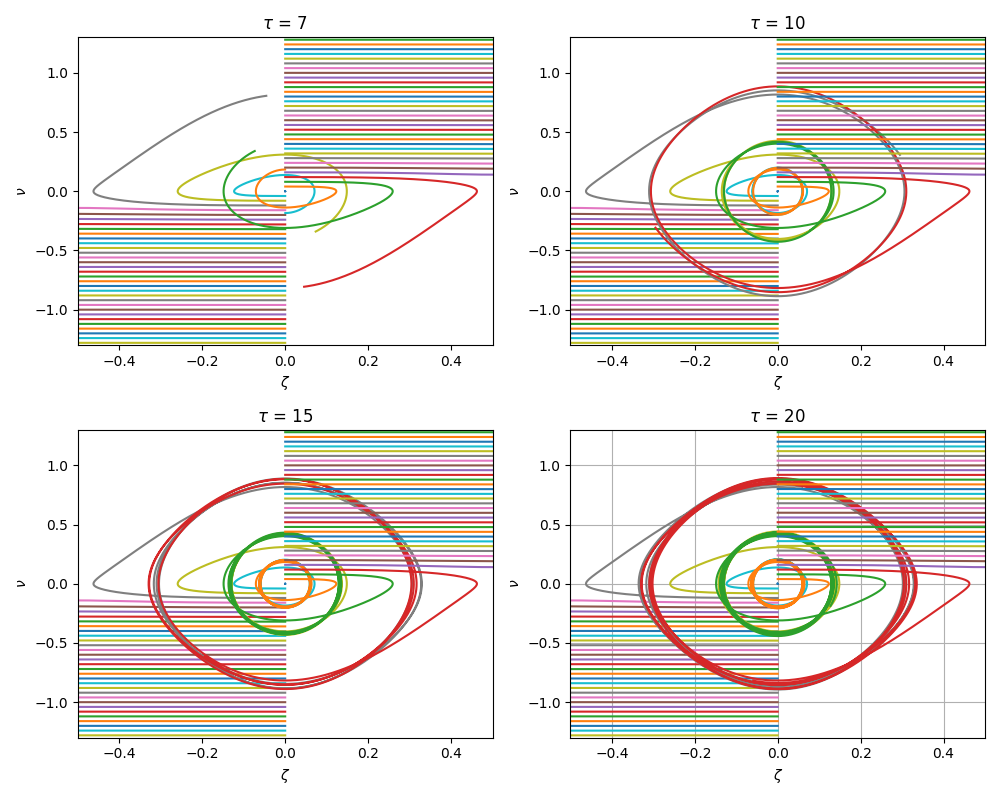}
    \caption{Formation of the island of trapped particles in phase space.}
    \label{fig:posvar}
    \end{figure}

The dynamics of the particles in phase space can also be visualized in Fig.~\ref{fig:posvar}. Note that, for larger and larger times, more and more passing particles become trapped, forming closed trajectories around the origin of the phase space. The frequency of this bounce motion is identified as the same frequency of the nonlinear oscillations of the BPI amplitude of Fig.~\ref{fig:el_field}. This result serves as a basis for the study of the next sections, where a beam of energetic ions nonlinearly interacts with the electric field of an EGAM instability in magnetized tokamak plasmas. 

\section{Gyrokinetic Model, equilibrium, and linear dynamics for EGAMs}
\label{sec:model}
The evolution of EGAMs is studied using ORB5, a global gyrokinetic code, see \cite{ORB5}, with collisionless electrostatic simulations. The model equations are obtained defining the phase space variables $ (\mathbf{R}, p_{||}, \mu )$, that are respectively the position, parallel momentum and magnetic momentum. The equations are the evolution in time of the markers coupled with the gyrokinetic Poisson equation, following the model given in Ref.~\cite{ORB5} (see also Ref.~\cite{Bottino2015} for a mathematical description of the electrostatic sub-model).

The tokamak magnetic equilibrium  considered here is the same as Ref.~\cite{Biancalani2017} and \cite{Biancalani2018}. The major radius is $R_0 = 1 \,m$, the minor radius is $a = 0.3125 \, m$, and we assume circular concentric  flux surfaces and a magnetic field on axis of $B_0 = 1.9 \,T$. The safety factor is uniform in space, equal at q = 2. The initial distribution function for the EPs is a double bump on tail. 

EGAMs grow unstable in this configuration due to the inverse Landau damping. When a scan with the EP concentration $n_{EP}$ is performed, one can see that this EGAM has a real frequency stemming from the GAM frequency at $n_{EP}=0$, and its frequency decreases with increasing $n_{EP}$. The growth rate increases from a negative value when $n_{EP}=0$ (corresponding to the GAM Landau damping), passing through zero (marginal stability) around $n_{EP}/n_i=0.05$, and growing with $n_{EP}$. Detailed studies can be found in Ref.~\cite{Biancalani2017}. 

\section{Study of the nonlinear oscillations of the EGAM amplitude}
\label{sec:nonlinear}

\begin{figure}[t!]
    \centering
    \begin{subfigure}{0.45\textwidth}
        \centering
        \includegraphics[width=\linewidth]{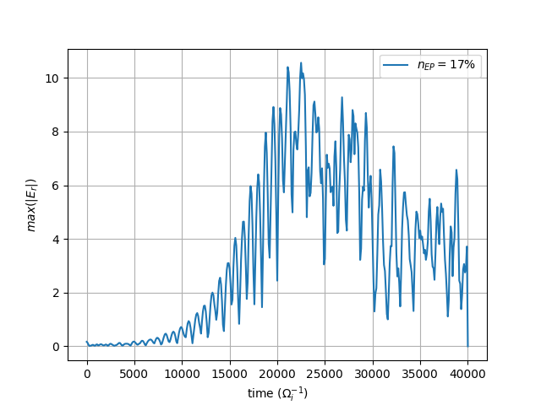}
        \caption{Maximum of the radial electric field as a function of time}\label{fig:EGAM_nEP17}
    \end{subfigure}
    \hfill
    \begin{subfigure}{0.45\textwidth}
        \centering
        \includegraphics[width=\linewidth]{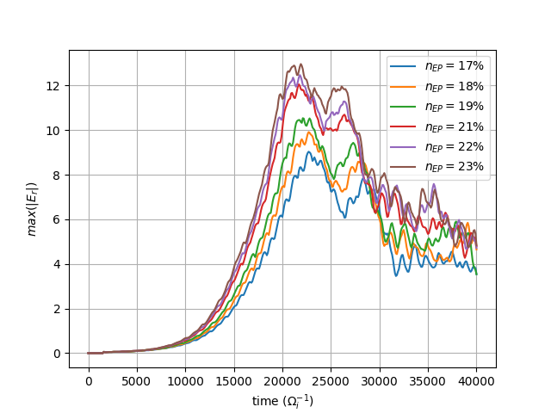}
        \caption{Mobile mean of the maximum at different EP concentration}\label{fig:EGAM_scan_nEP}
    \end{subfigure}
    \label{EGAM}
\end{figure}

The EGAM linear and nonlinear dynamics can be investigated by measuring the evolution of the radial electric field in time. In Fig.~\ref{fig:EGAM_nEP17}, the case of a simulation with $n_{EP}/n_i=0.17$ is considered. One can observe a linear phase, lasting from the beginning of the simulation until $t\simeq 20000 \; \Omega_i^{-1}$, a saturation at  $t\simeq 22000 \; \Omega_i^{-1}$, and a decrease in amplitude. The signal shows the co-existence of two scales: a high frequency component, which is the EGAM real part of the frequency, and a lower frequency component, after the nonlinear saturation. By averaging out the high-frequency part, we are left with the absolute value of the EGAM amplitude. 

\begin{figure}[b!]
    \centering
    \includegraphics[width=0.5\textwidth]{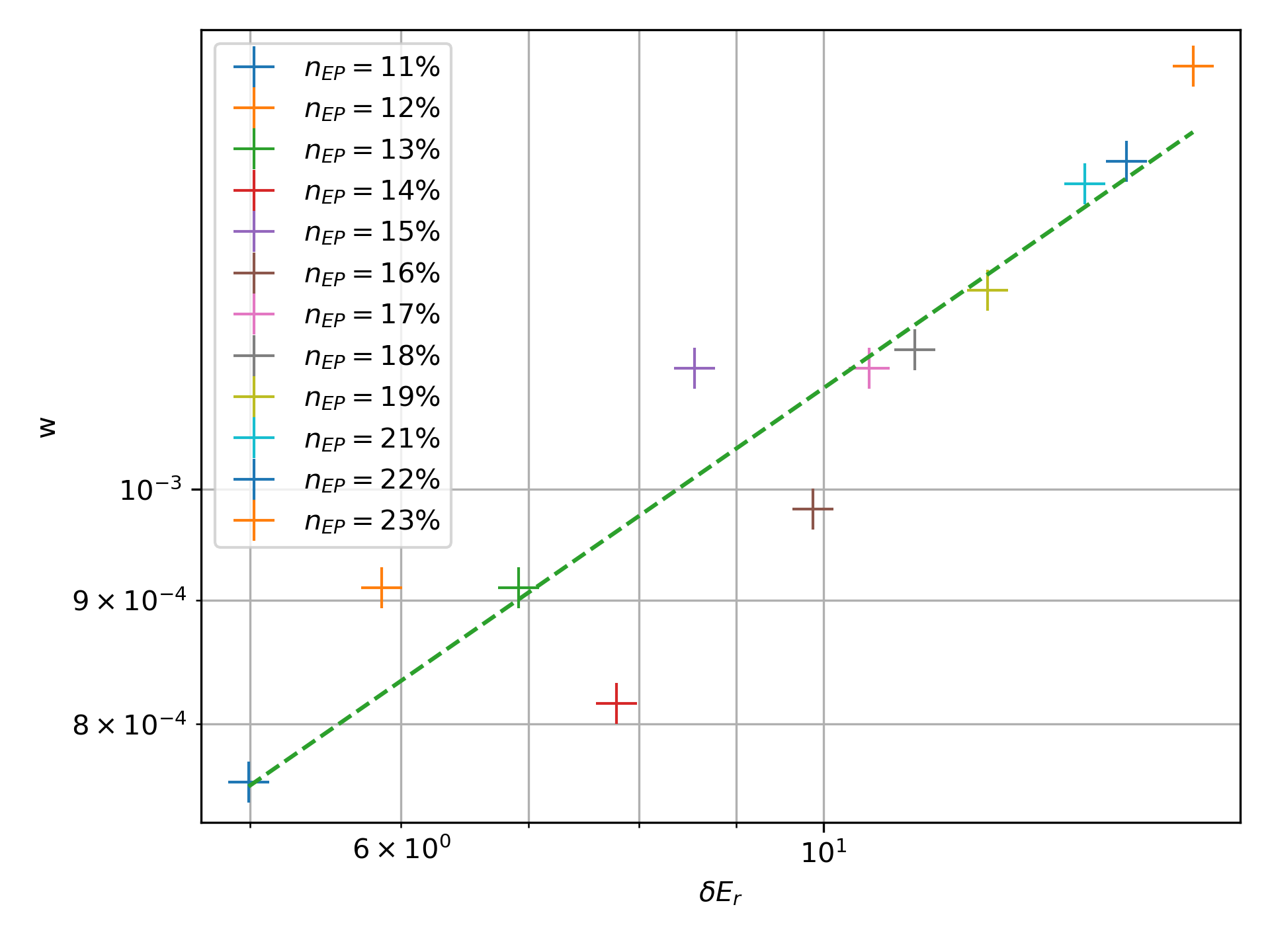}
    \caption{nonlinear oscillations as a function of the saturation level at different EP concentrations on a logarithmic scale}
    \label{scan}
\end{figure}

The evolution of the EGAM amplitude, for different values of EP concentration, is shown in Fig.~\ref{fig:EGAM_scan_nEP}. The concentration of energetic particles has been varied from 11\% to 22\%. Note that a clear low-frequency oscillation of the EGAM amplitude is observed in the nonlinear phase after the saturation, for at least one of two periods, then fading into noise. We measure this nonlinear frequency, and plot it versus the EGAM saturation level $\delta E_r$. The result can be seen in Fig.~\ref{scan}, on a logarithmic scale. Doing the linear fitting of the points we find the scaling of the frequency of the nonlinear oscillations $\omega_{nl}$:
\begin{equation}\label{eq:omega_nl_vs_deltaEr}
\omega_{nl} \propto {\delta E_{r}}^{\alpha}    
\end{equation}
For the regime considered here, and with the scan used, we find a value of $\alpha\simeq 0.6$.

Note that the saturation level of EGAMs has been described in previous papers, and its scaling with the linear properties of the EGAM has been given. Therefore, in principle, thanks to Eq.~\ref{eq:omega_nl_vs_deltaEr}, we provide a way to predict the frequency of the nonlinear oscillations of EGAMs, given the linear properties. This can be used in the framework, for example, of quasilinear models, which need reduced models of the nonlinear properties, as function of the linear properties.

\section{Comparison with BPI}
\label{sec:comparison}

As already shown, there's a one-to-one correspondence between the Beam Plasma Instability and the saturation mechanism of EGAMs. In Fig.~\ref{fig:el_field} we can see the evolution of the amplitude of the Langmuir wave, obtained by solving the model proposed in Ref.~\cite{Levin72}. In his model the frequency of the wave is supposed to be fixed, so that we do not see the frequency evolution we can see in the EGAMs. The saturation mechanism is similar: after a linear phase the wave saturates due to the energetic particle redistribution in phase space. We can see here the nonlinear oscillations that we analyzed in the previous section. In this case, as measured by Levin the frequency of the oscillation is of the same order of the bounce frequency of the particles.

In this section, we want to investigate if such a dependence also occurs for EGAMs. To this aim, we are interested in finding a possible scaling between the EGAM nonlinear frequency and the bounce frequency of trapped particles, as shown for BPI by Levin in 1972, see Ref.~\cite{Levin72}. 
The nonlinear EGAM frequency can be measured by detecting numerically the peaks of the low-frequency oscillations. On the other end we use the result obtained by Qiu in 2011(\cite{Qiu2011}), and verified  by Biancalani in 2018(\cite{Biancalani2018}) that links the bounce frequency with the saturation value. The equation is the following:
\begin{equation}
    \omega_b^2 = \frac{e\hat{V}_{dc}}{2m_{EP}v_{||0}qR_0} \delta E_r.
    \label{bounce}
\end{equation}

The mass of energetic particles, $m_{EP}$, is considered to be the mass of the bulk, $v_{||0} $ is the resonant velocity and $\hat{V}_{dc} = \frac{m_{EP}v_{||0}^2}{eBR}$ is the magnetic curvature drift. 
So we can compare the nonlinear oscillations with the saturation level of the electric field.

By using eq. (3) and (4) we can see that even in EGAMs the nonlinear frequency and the bounce frequency are linked by the relation:
\begin{equation}
\omega_b \propto \omega_{nl}^{\beta}
\end{equation} 
with $\beta=1/2\alpha$. In the regime considered, here, we obtain $\beta_{EGAM}\simeq 0.83$, which is very similar (within the error bars of the measurements of our GK simulations) to the value of the BPI: $\beta_{BPI}\simeq 1$. This result shows one more similarity between the EGAM and the BPI. This similarity can be used to better understand the physics of the EGAM nonlinear dynamics, by using reduced models such as a solver for the BPI problem.

\section{Summary and Outlooks}
\label{sec:conclusion}

In this paper, we have compared the nonlinear dynamics of the beam-plasma instability (BPI) and the energetic particle induced geodesic acoustic mode (EGAM). EGAMs are zonal instabilities observed in tokamaks, driven unstable by the nonuniformity of energetic particles (EP). Note that perturbation of a zonal electric field, like the one of the EGAMs, can be studied as a renormalized nonlinear equilibrium, dubbed  Zonal State~\cite{Falessi19,Falessi23,Qiu25}.

The BPI amplitude is studied adopting the model of Levin~\cite{Levin72}, where the Langmuir wave frequency and radial structure are approximated as constant, and only the amplitude is let evolve in time, in a self-consistent way with respect to the electron beam dynamics. The nonlinear saturation of this instability occurs due to the change in trajectory of the electrons, going from passing to trapped in the wave. Nonlinear oscillations of the BPI amplitude are observed and linked to the bounce frequency of the trapped electrons. This result paves the way for the second part of this study, where we consider EGAMs in tokamak plasmas.

We study the EGAMs by means of global gyrokinetic simulations using the particle in cell code ORB5~\cite{ORB5}. We find that the EGAM grows linearly, then saturates due to a similar mechanism as the BPI: a redistribution of the EPs in phase space. We observe nonlinear oscillations of the EGAM electric field, right after the saturation. The study of this nonlinear frequency, for different values of the drive, allows us to find a similar linear scaling of the nonlinear frequency with the bounce frequency of the particles trapped in the wave. The nonlinear frequency is found to scale with a power of $0.6$ as a function of the field, which is very similar (within the measurements error bars) to the value expected for the BPI (which goes like the square root of the field amplitude). This confirms that the nonlinear dynamics of forced oscillations like EGAMs in tokamak plasmas, has strong similarities with the BPI in uniform plasmas. Both saturate due to the EP redistribution in phace space.

Following this result, we propose here a novel diagnostics for the measurement of the EGAM amplitude in tokamak plasmas. This electric field is often difficult to measure due to the lack of diagnostics capable of entering the inner part of the tokamak. Here we propose to measure the frequency of the nonlinear oscillations of the field amplitude, instead. This can be found with diagnostics which are positioned outside the tokamak. Once this frequency is measured, the field amplitude can be indirectly estimated using the scaling we provide in this paper.

\section*{Acknowledgments}
Discussions with Ph. Lauber, Th. Hayward-Schneider, A. Mishchenko, D. Gossard, and R. Ivanov are gratefully acknowledged.
This work was performed with the ORB5 code on the Pitagora Supercomputer at CINECA.
This work has been carried out within the framework of the EUROfusion Consortium, partially funded
by the European Union via the Euratom Research and Training Programme (Grant Agreement No
101052200 — EUROfusion).

\printbibliography
\end{document}